\documentclass[twocolumn,tighten]{aastex61}
\hypersetup{linkcolor=blue,citecolor=blue,filecolor=cyan,urlcolor=blue}

\def\arcdeg{\hbox{$^\circ$}}
\def\arcmin{\hbox{$^\prime$}}
\def\arcsec{\hbox{$^{\prime\prime}$}}

\def\fdg{\hbox{$.\!\!^\circ$}}
\def\farcm{\hbox{$.\mkern-4mu^\prime$}}
\def\farcs{\hbox{$.\!\!^{\prime\prime}$}}

\def\nodata{ ~$\cdots$~ }

\usepackage{graphicx}
\usepackage{rotating}
\usepackage{longtable}

\shorttitle{Proper motions in ONC}
\shortauthors{PLATAIS ET AL.}

\begin{document}

\title{{\it HST} astrometry in the Orion Nebula Cluster: census of low-mass runaways}

\correspondingauthor{Imants Platais}
\email{imants@jhu.edu}

\author{Imants Platais}\affil{Department of Physics and Astronomy, Johns Hopkins University, 3400 North Charles Street, Baltimore, MD 21218, USA}

\author{Massimo Robberto}\affil{Space Telescope Science Institute, 3700 San Martin Drive, Baltimore, MD 21218, USA}

\author[0000-0003-3858-637X]{Andrea Bellini}\affil{Space Telescope Science Institute, 3700 San Martin Drive, Baltimore, MD 21218, USA}

\author{Vera Kozhurina-Platais}\affil{Space Telescope Science Institute, 3700 San Martin Drive, Baltimore, MD 21218, USA}

\author{Mario Gennaro}\affil{Space Telescope Science Institute, 3700 San Martin Drive, Baltimore, MD 21218, USA}

\author{Giovanni Strampelli}\affil{Space Telescope Science Institute, 3700 San Martin Drive, Baltimore, MD 21218, USA}

\author{Lynne A. Hillenbrand}\affil{Department of Astronomy, California Institute of Technology, Pasadena, CA 91125, USA}

\author{Selma E. de Mink}\affil{Center for Astrophysics, Harvard-Smithsonian, 60 Garden Street, Cambridge, MA 02138, USA}\affil{Anton Pannekoek Institute for Astronomy, University of Amsterdam, Science Park 904, 1098 XH, Amsterdam, The Netherlands}

\author{David R. Soderblom}\affil{Space Telescope Science Institute, 3700 San Martin Drive, Baltimore, MD 21218, USA}

\begin{abstract}

We present a catalog of high-precision proper motions in the Orion
Nebula Cluster (ONC), based on Treasury Program observations with the
{\it Hubble Space Telescope's} ({\it HST}) ACS/WFC camera.  Our
catalog contains 2,454 objects in the magnitude range of $14.2<m_{\rm
  F775W}<24.7$, thus probing the stellar masses of the ONC from
$\sim$$0.4 M_\sun$ down to $\sim$$0.02 M_\sun$ over an area of
$\sim$550 arcmin$^2$. We provide a number of internal velocity
dispersion estimates for the ONC that indicate a weak dependence on
the stellar location and mass. There is good agreement with the
published velocity dispersion estimates, although nearly all of them
(including ours at $\sigma_{v,x}=0.94$ and $\sigma_{v,y}=1.25$
mas~yr$^{-1}$) might be biased by the overlapping young stellar
populations of Orion\,A. We identified 4 new ONC candidate runaways
based on {\it HST} and the {\it Gaia} DR\,2 data, all with masses less
than $\sim$1\,$M_\sun$. The total census of known candidate runaway
sources is 10 -- one of the largest samples ever found in any Milky
Way open star cluster. Surprisingly, none of them has the tangential
velocity exceeding 20~km~s$^{-1}$. If most of them indeed originated
in the ONC, it may compel re-examination of dynamical processes in
very young star clusters. It appears that the mass function of the ONC
is not significantly affected by the lost runaways.

\end{abstract}


\section{Introduction}
\label{intro}

As one of the most recognizable objects in the astronomical sky and
one of the most popular in studies of star formation, the Orion Nebula
Cluster (ONC) may not need a formal introduction. Its basic properties
are reviewed by, e.g., \citet{od01,od08,mu08}. Over the past two
decades, our understanding of the ONC has significantly improved, as
evidenced by several hundred publications listed in
SIMBAD\footnote{{\texttt http://www.simbad.harward.edu/simbad/}}
astronomical bibliography.  Our contribution to this huge body of
various data for the ONC is a new set of relative proper motions
obtained from a massive imaging effort by {\it HST}.

Until recently, there were only two studies \citep{jo88,va88} that
provided accurate relative proper motions, albeit limited to stars
brighter than $I\leq$16 magnitude ($V\sim20$). Both studies produced a
clean sample of cluster members indicated by high membership
probabilities: $P_{\mu}>90\%$. The calculated proper-motion dispersion
along one axis ranges from 0.76 mas~yr$^{-1}$ to 1.18 mas~yr$^{-1}$.
Assuming the ONC distance to be 414$\pm$7~pc \citep{me07}, this range
of dispersions translates into the velocity dispersions of 1.5 and
2.3~km~s$^{-1}$.  Both studies indicate a larger velocity dispersion
in the $Y$ direction (South-North).

More recent contributions to the kinematics of ONC members are the
studies by \citet{ku19} and \citet{ki19}. The first of these studies
analyzed {\it Gaia} DR\,2 \citep{gaia2} proper motions in 28 young
star clusters and associations.  Among them, the ONC had special
attention due to its prominent role in our understanding of star
formation. These authors found that the ONC is an ordinary
gravitationally-bound cluster, while the known asymmetry in the
internal velocity distribution is re-confirmed. Similar conclusions
were reached by \citet{ki19} using archival ONC images from various
{\it HST} cameras and from ground-based Keck NIRC2 data, over a small
field centered onto the Trapezium.

These new proper-motion measurements have also stimulated searches of
the objects likely ejected from the ONC \citep{ki19,mk19}, and
generated lists of potential low-mass runaways
($<1\,M_\sun$). \citet{wa19} performed $N$-body simulations,
specifically designed to imitate the ONC, which indicate that, within
1~Myr, a few-body dynamical decay can eject a few massive OB stars.
These simulations, however, do not address the low-mass escapees.
More universal appear to be the $N$-body simulations by \citet{mo13}
which consider a variety of dynamical interactions between the stars
and potential mechanisms of making the runaways at a large range of
escape velocities and stellar masses. Over 4~Gyr, this study predicts
a high percentage of slow-moving runaways.

The supremacy of {\it Gaia} absolute astrometry appears to be
unassailable.  Then, what is the contribution of our new relative
proper motions?  The strongest argument is that the {\it Gaia}
limiting magnitude is at $G\sim21$~mag, while {\it HST} can observe
objects several magnitudes fainter. Another issue is the
highly-irregular nebulosity that forms a backdrop to the ONC and
presumably adds semi-random noise to the {\it Gaia} measurements,
effectively lowering the spatial resolution.  In addition, the ONC
contains objects whose apparent profile significantly differs from
that of the stars (e.g., protoplanetary disks dubbed as proplyds).  As
a result, such objects produce a poor fit by a standard stellar
point-spread function (PSF). For such objects, we could utilize a
template-fitting method developed specifically for {\it HST} images by
\citet{ma08}, but in practice it is not feasible due to the low number
of images. In general, the precision of a single positional
measurement with {\it HST} is on a par with that of {\it Gaia}. All
these reasons make the {\it HST} dataset on the ONC competitive or
even superior over small spatial fields.

We concentrate on the kinematic properties of ONC members via a new
survey of proper motions. Astrometric measurements in this area are
challenging.  Therefore, we provide a detailed account on how to get
from centroids of objects to the catalog of proper motions
(Section~\ref{hstobs}) and how they can be interpreted
(Section~\ref{discuss}). Our conclusions can be found in
Section~\ref{conclu}.

\section{{\it HST} Surveys of Orion Nebula Cluster and Data
Reductions}\label{hstobs}

The Treasury Program on the ONC (GO-10246; PI: M. Robberto) produced a
large number of observations with three different {\it HST} cameras
through 10 different filters \citep{ro13}. The main purpose of this
program was to obtain photometry in various bandpasses over a
contiguous area of the sky in order to characterize all detected
sources such as stars, circumstellar disks, proplyds and brown
dwarfs. In the region of the ONC, the spatial density of sources is
intrinsically low. If we consider the {\it HST} imaging instrument
with the largest field-of-view (FOV) covering $\sim$11 arcmin$^2$ --
the Wide-Field Channel of the Advanced Camera for Surveys (ACS/WFC) --
then even in the Trapezium area a single long-exposure can only detect
$\sim$300 sources with a signal-to-noise ratio above 5.  The number of
detected sources quickly drops to $\sim$50 or fewer outside the
Trapezium. Many of these sources are not appropriate for
high-precision astrometry (especially proplyds), and that further
reduces the number of available ``astrometric'' stars.

This first {\it HST} Treasury Program on the ONC was done over two
sets of epochs: 2004 October~11 - November~7, and 2005 March~3 -
April~26, with a $\sim$180$\arcdeg$ change of the {\it HST}
orientation between these periods. These two sets of observations
constitute our first epoch.  In order to achieve complete spatial
coverage with two cameras, ACS/WFC and WFPC2 (the Wide-Field and
Planetary Camera~2), the observations were acquired along nine strips
oriented close to the East-West direction so that, for the ACS/WFC,
there is a $\sim$50\% overlap between adjacent pointings in this
direction.  The drawback of this observing strategy is a small and
variable 5-40\% overlap in the South-North direction \citep[see
  Figure~2 in][]{ro13}.  Such a limitation is undesirable for the
construction of astrometric reference frame, if we have to rely
entirely on {\it HST} observations \citep[e.g.,][]{pl15}.

A decade later, we designed a second {\it HST} Treasury Program on the
ONC (GO-13826; PI: M. Robberto) that addresses some of the
shortcomings listed above. For the purpose of deriving proper motions,
the ACS/WFC observations were designed to reproduce -- to the extent
possible -- the same pointings and exposure time, telescope
orientation, guide stars, and the Earth-orbit position as in our first
survey. In order to optimize the telescope time, we used only the
primary pointings (visits, a total of 52) in program GO-10246 and
ignored all the 50\%-offset pointings. The necessary overlap between
adjacent pointings and strips is achieved by adopting larger dithers.
Only the broadband filter F775W is used in these observations over two
times in 2015 February~8 - April~28, and 2015 August~1 - October~29,
thus establishing our second epoch. We note that, due to the problems
related to finding appropriate guide stars, four visits have a
significant spatial offset with respect to the corresponding GO-10246
first-epoch pointings, and 17 visits have a slightly different roll
angle.

In what follows, we used only the ACS/WFC images taken through filters
F775W and F555W. These ACS/WFC filters are well-calibrated
astrometrically, provide a similar saturation level and limiting
magnitude, and sample the ONC area equally in both Treasury
Programs. We employed all available 340 to 385\,s long exposures, and
also all those 8\,s exposures that contain a reasonably high number of
stars. This amounted to 196 first-epoch images and 259 second-epoch
images. From the Mikulski Archive for Space telescopes we downloaded
the bias-subtracted, dark-subtracted, flat-fielded, and corrected for
charge-transfer inefficiency \texttt{\_flc.fits} files.  We used the
software code \texttt{img2xym\_WFC.09x10} \citep{an06} to calculate
precise positions and instrumental magnitudes for all detectable
sources, and a quality-of-fit (\texttt{qfit}) parameter.

\begin{figure}
{\includegraphics[width=\columnwidth]{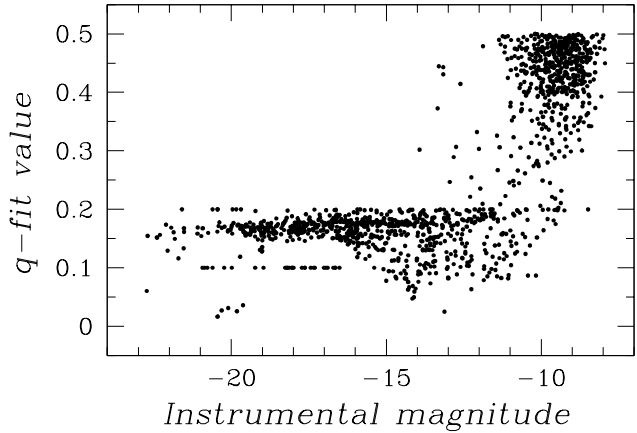}}
\caption{Distribution of the quality parameter \texttt{qfit} for sources
measured in the central ONC image \texttt{jcol35inq}. Only a small fraction
of detections with instrumental magnitudes brighter than $-16$ are stars and
 other sources. }
\label{fig:qfit}
\end{figure}

\subsection{Geometric Distortion Corrections}\label{geocor}

A standard reference for the geometric distortion of the ACS/WFC is
provided by \citet{an07}, with additional improvements provided by
\citet{ub13}.  However, a recent astrometric study in the region of 30
Dor \citep{pl15} indicates that the existing corrections can be
further improved. This prompted us to obtain a new set of corrections
based on the extensive new study of the geometric distortion of the
ACS/WFC \citep{ko15}. The principal difference between the previous
work by \citet{an07} and \citet{ko15} is the separation of the
fine-scale distortion of the detector (lithographic-mask pattern) from
the filter-dependent distortion in the latter study.  Still, the study
by \citet{ko15} is limited by the accuracy of the original {\it HST}
standard astrometric catalogs. This issue has been addressed by
\citet{koz18} using the {\it Gaia} Data Release~1
\citep[DR\,1;][]{gaia}. The authors show that the original {\it HST}
standard catalogs are affected by measurable differences in rotation
and scale with respect to the {\it Gaia} DR\,1 and also contain a
small uncorrected skew term.  We implemented the necessary upgrades to
the existing astrometric standard for the globular cluster 47~Tuc so
that it is now on the system of {\it Gaia} DR\,1.  Then the improved
astrometric standard catalog was used to re-calibrate the distortion
solutions for the ACS/WFC F555W and F775W filters at both epochs of
our ONC observations.  With the new set of constants and look-up
tables accounting for a total of four separate components in the
ACS/WFC geometric distortion (correlated detector-grid imperfections,
tiny filter flaws, polynomial part of distortion, and the
time-dependent skew correction), we expect our measured positions of a
single image to be accurate at the 2-3 mas.  For the ONC region, this
is crucial because the small number of reference stars per image
prevents us from using the so-called local solutions in calculating
proper motions which are largery immune against the imperfect
geometric distortion corrections.

\begin{figure}
{\includegraphics[width=\columnwidth]{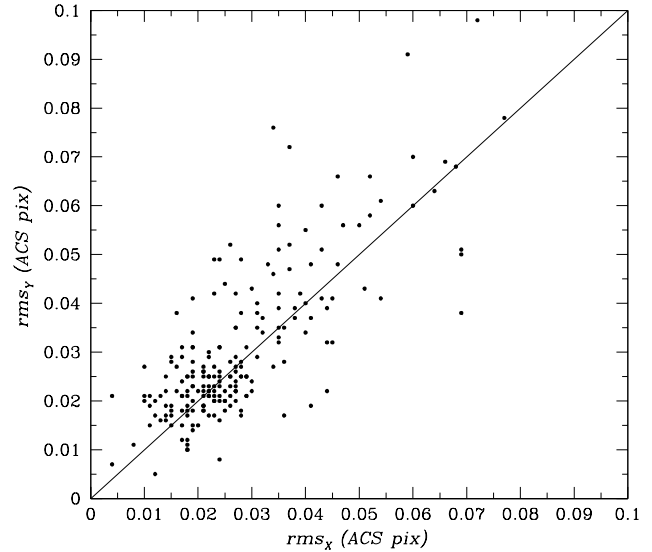}}
\caption{Distribution of rms errors from the least-squares transformations for
all frame-tile, tile-strip, and strip-strip solutions. Nomenclature 
of these solutions is provided in \citet{pl15}. The small number of
common stars in several frame-tile solutions usually yields a high rms error.}
\label{fig:rms_all}
\end{figure}

\subsection{Setting up the Astrometric Reference Frame}\label{frame}

The sky area in the direction of the ONC is very complex: a bright and
patchy nebulosity, the presence of bright stars, and the intrinsically
low number of stars. Such conditions are not favorable to
high-accuracy astrometry with {\it HST} over the small FOV of its
imaging instruments.  {\it Gaia} is expected to provide a major
improvement to the absolute astrometry in this area. Nevertheless, the
{\it HST} contribution is essential at fainter magnitudes, beyond the
{\it Gaia} detection limit.  For this, we need to construct an
astrometric reference frame that serves as a ``touchstone'' to our
selected 455 ACS/WFC images of the ONC.

Our initial choice was to base the reference frame on the VISTA
Orion\,A Survey \citep{me16}, a deep near-infrared survey that
contains $\sim$40 counterparts per single ACS/WFC image within our
FOV. The epoch of the VISTA survey near the ONC is 2013.2, which is
reasonably close to our second epoch. This makes the VISTA Survey a
suitable candidate for the astrometric reference frame, once the
positions of stars are translated into the {\it Gaia} system. Then,
each second-epoch ACS/WFC frame can be transformed into the revised
VISTA coordinates by using least-squares and a polynomial model. The
typical residuals of this coordinate-transformation are on the order
of $\sim$20~mas. Next, we followed closely the procedures to compute
proper motions, described in detail by \citet{pl18}. However, the
resulting preliminary proper motions of the ONC objects showed a
$\sim$20\% larger internal velocity dispersion than the existing
studies \citep[e.g.,][]{jo88}.  Apparently, the existing ground-based
observations over spatial scales exceeding $\sim$1$\arcmin$ may not
produce an adequate positional precision to support astrometry from
{\it HST} without a significant loss of accuracy.  Therefore, we had
no viable alternative but to construct an astrometric reference frame
directly from the {\it HST} observations, since the available {\it
  Gaia} high-quality astrometry around the ONC is too sparse.

\begin{figure}
{\includegraphics[width=\columnwidth]{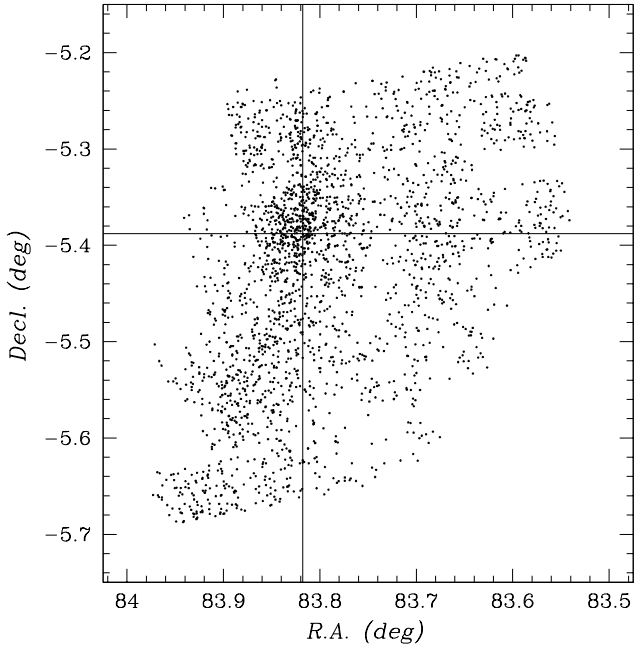}}
\caption{Spatial distribution of all sources with measured proper motions.
Location of the Trapezium is marked by crosshairs.}
\label{fig:spat_distr}
\end{figure}

\begin{figure}
{\includegraphics[width=\columnwidth]{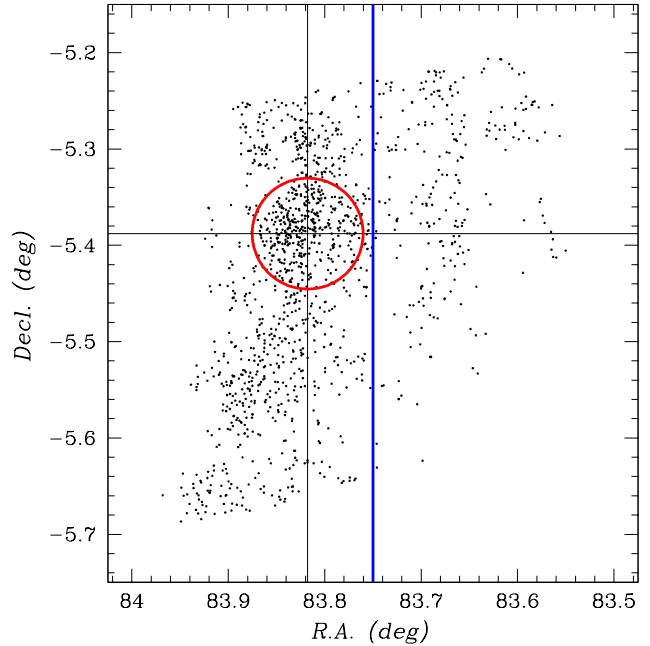}}
\caption{Spatial distribution of the selected 1379 sources with more
  reliable proper motions. Red circle: inner area of the ONC with a
  radius of $3\farcm45$ centered on Trapezium; vertical blue line:
  cut-off at R.A.$=83\fdg75$ separating the western sources with a
  possibly lower accuracy of proper motions. }
\label{fig:best_spat_distr}
\end{figure}

Once we have distortion-corrected positions with significant spatial
overlap, a linear three-term transformation (offset, rotation, scale)
puts one set of positions (a tile) into the system of
partially-overlapping adjacent tiles.  To contruct the reference
frame, only second-epoch observations and long exposures are
used. These observations are better designed in terms of astrometry
and also overlap chronologically with the {\it Gaia} measurements.
Following the hierarchical accumulation algorithm by \citet{pl15}, we
chose an image of the Trapezium area (\texttt{jcol35inq}) as the seed
of a global coordinate system for the ONC.  In practice, a combination
of heavy contamination from ``bleeding'' sources, numerous stellar
impostors (see Figure~\ref{fig:qfit}) and a limited number of common
stars between adjacent tiles (frequently less than 10), made some
transformations possible only interactively. The distribution of the
rms errors for all transformations is shown in
Figure~\ref{fig:rms_all}.  The majority of rms errors are concentrated
at $\sim$0.02 pixel, equivalent to 1~mas on the sky. This is
comparable to the rms scatter listed for the ACS/WFC observations in
the 30 Doradus region \citep[Table~2;][]{pl15}, despite a huge
difference in the number of available common stars.  None of our
transformations is based on more than 60 stars.  As the result, we
obtained a catalog of 10,409 sources in our area of the ONC, that is a
mixture of real stars, other celestial objects, and all kinds of
spurious objects. In order to align the catalog along the R.A. and
decl. axes, it is rotated by $82\fdg2309$. These rotated $XY$
coordinates provides our astrometric reference frame in units of the
ACS/WFC distortion-corrected pixels.

In order to calculate equatorial coordinates of this astrometric reference
frame, we used 452 stars from {\it Gaia} DR\,2  and applied a linear
3-parameter polynomial model in each coordinate.  The resulting residuals
indicate semi-gradual  offsets up to $\sim$$0\farcs07$ across the FOV,
equivalent to
$\sim$1.4 ACS/WFC pixel. Such a pattern and the amplitude of these offsets
were also noticed by \citet{pl18}. In spite of our significantly improved
corrections for geometric distortions, the nature of these offsets for
the ACS/WFC is still an open question. We believe that the final {\it Gaia}
astrometric catalog will help to eliminate the issue. In the following
we argue that the imperfections of astrometric reference frame
are not critical to our scientific results and their interpretation.
If some portion of the final data catalog appears to be suspicious, then
it is not used in our analysis, or are marked in our catalog.

\subsection{Proper Motions and Positions}\label{promo}

The next step towards calculating proper motions is transforming all
sets of pixel coordinates into the system of astrometric reference
frame \citep{pl15}. A linear 3-parameter polynomial model in $X$ and
$Y$ is used in the least-squares transformations. A notable difference
between \citet{pl15} and our study is that we used combined sets of
input coordinates, while in \citet{pl15} each ACS/WFC chip was
transformed separately. This is also a default mode in applying the
corrections for geometric distortion and is preferable when the number
of stars is low. For our second-epoch observations, the solution's rms
error is very small -- at a level of 0.014 ACS/WFC pixel. However,
similar first-epoch transformations produce much higher rms scatter --
on average, 0.21 ACS/WFC pixel -- due to the effect of proper motions
over $\sim$10 years. There are two potential issues with this. First,
such a large scatter acts as a source of additional noise and,
therefore, sets a limit to the final precision and accuracy of our
relative proper motions.  Second, in the case of a very low number of
reference stars, $4 \le\,n\,\le 10$, that may bias the resulting
proper motions, effectively reducing them.  Fortunately, the pattern
of our second-epoch pointings partially mitigate the impact of these
issues.  We note that the {\it Gaia} proper motions are immune to such
effects, albeit they are still affected by nebulosity in the direction
of the ONC.

\begin{figure}
%\resizebox{\hsize}{!}
{\includegraphics[scale=0.75]{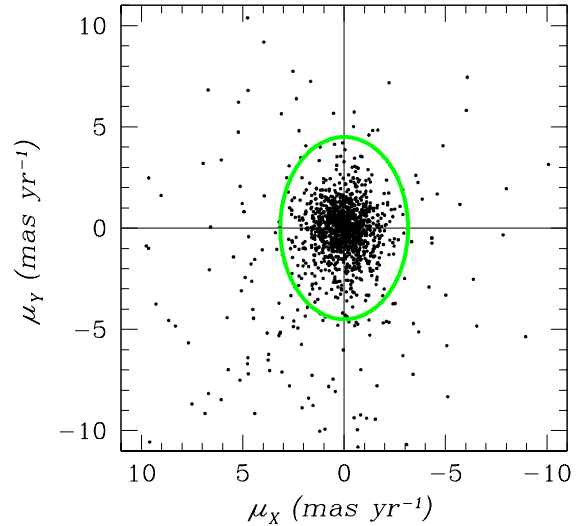}}
\caption{Vector-point diagram of sources shown in
Figure~\ref{fig:best_spat_distr}. Green ellipse indicates the limit of
selected proper motions used to calculate the internal velocity dispersion of
the ONC. 
}
\label{fig:best_vpd}
\end{figure}

\begin{figure*}
%\resizebox{\hsize}{!}
\centering
{\includegraphics[width=14cm]{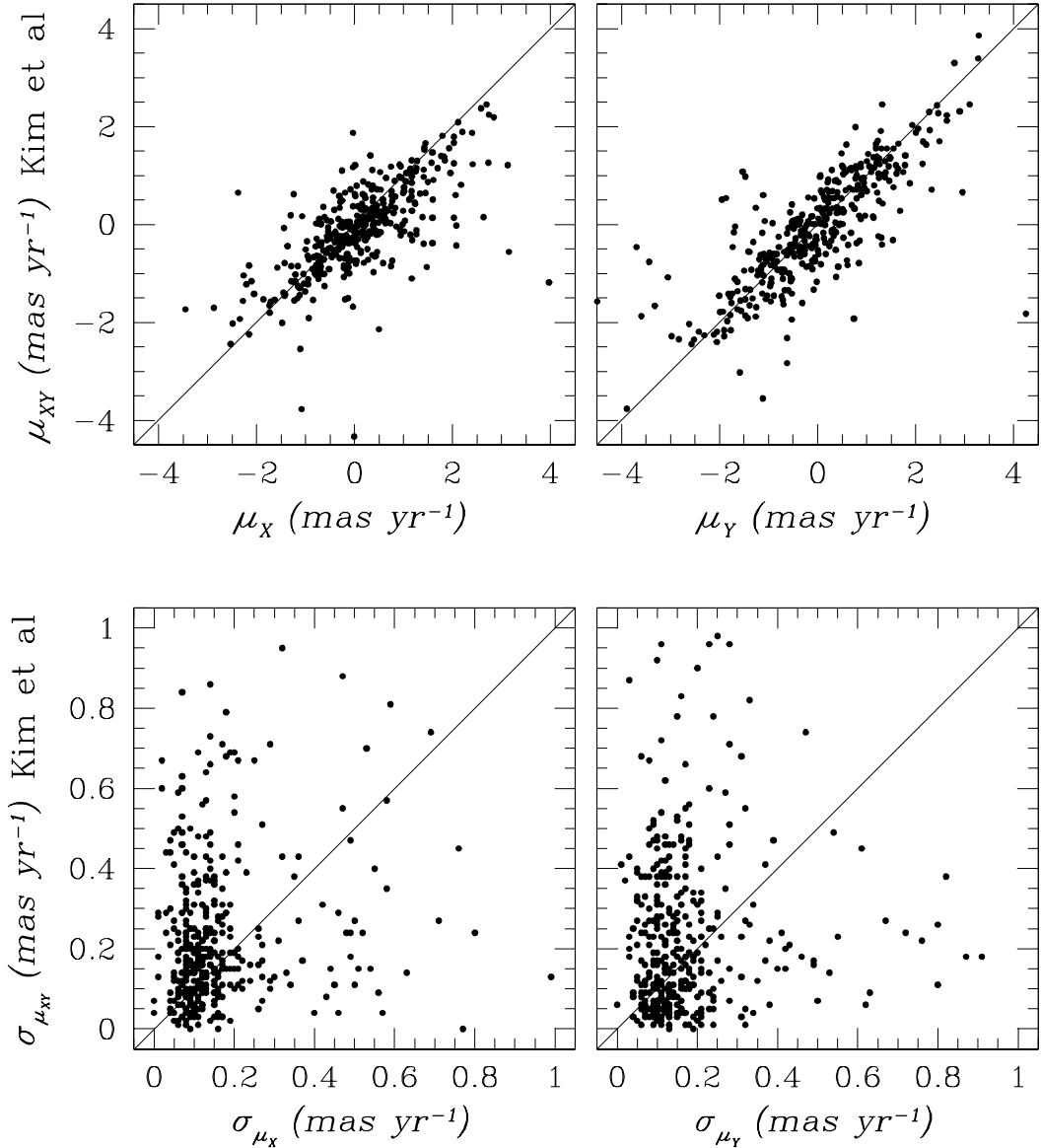}}
\caption{Comparison of proper motions and their errors between this study
and that by \citet{ki19}. While the proper motions are distinctly
commensurated, the corresponding distributions of proper-motion errors are
lopsided.
}
\label{fig:kim_plat}
\end{figure*}

The total number of first-epoch detections in the system of our
astrometric reference frame is 42,000, while there are 136,000
second-epoch detections. The majority of the latter are hot pixels and
cosmic rays.  The detections in both epochs contain a large number of
other artifacts such as ``granulated'' spikes of the brighter stars,
especially near the Trapezium. In fact, the real detections are so
polluted by these artifacts that we could not construct a reasonable
global master list \citep[such as in Sect.~3.3.1,][]{pl15}. Therefore,
we adopted the ACS source catalog \citep[Table~5,][]{ro13} as a clean
and complete master list.  The original equatorial coordinates of this
catalog were translated into the {\it Gaia} DR\,2 system by using the
nearest 9 common stars and their R.A. and decl. offsets.  Then, the
updated celestial coordinates were translated into the system of our
astrometric reference frame.

Proper motions are calculated following the scheme outlined in
\citet{pl15}.  Around each entry in the master list, we selected all
detections (subsets) within a radius of 9 ACS/WFC pixels, equivalent
to 450 mas.  This size was chosen in order to find large proper
motions in this area of the sky. On the other hand, it may be too
large for calculating proper motion in the case of visual binaries and
their components.  Each detection has its estimated standard error
based on the instrumental magnitude, thus allowing us to perform the
weighted least-squares fit to the measurements in each coordinate as a
function of time.  We rejected the most deviant measurement, if its
offset is larger than 4$\sigma$. This extirpation is repeated, if
necessary, until the lowest limit of measurements ($n$=3) is reached
\citep{pl15}. If an epoch has only two detections, then none of them
is deleted. We caution that, in some of such cases, the resulting
large proper motion might be spurious. A typical hint of such a
failure can be a minimum number of datapoints in combination with
significantly higher proper-motion errors. Our proper-motion catalog
(Table~\ref{tab:catal}) contains 2,454 objects in the magnitude range
of $14.2<m_{\rm F775W}<24.7$.  The highest average precision of proper
motions at 0.11 mas~yr$^{-1}$ is obtained for objects in the magnitude
range of $18.0<m_{\rm F775W}<21.0$ mag.

\begin{deluxetable*}{lcl}
 \tabletypesize{\footnotesize}
 \tablecolumns{3}
 \tablewidth{0pt}
 \tablecaption{Proper Motion Catalog\label{tab:catal}}
 \tablehead{
    \colhead{Unit}  &
    \colhead{Label} &
    \colhead{Explanations}
    }
\startdata
 ---    &   ID     & Number from Robberto et al. (2013) \\
 mas~yr$^{-1}$ &   pmx    & weighed proper motion in X \\
 mas~yr$^{-1}$ &   pmy    & weighed proper motion in Y \\
 mas~yr$^{-1}$ & e\_pmx     & error of the weighed proper motion in X \\
 mas~yr$^{-1}$ & e\_pmy     & error of the weighed proper motion in Y \\
 ---    &   cx   & normalized $\chi^2$ for proper motion in X\\
 ---    &   cy   & normalized $\chi^2$ for proper motion in Y\\
 ---    &   qx    & goodness-of-fit probability Q in X\\
 ---    &   qy    & goodness-of-fit probability Q in Y\\
 mag    &   F775W  & preliminary F775W magnitude \\
 yr     &   ep\_e  & maximum extent of epochs \\
 ---    &   n1     & number of first-epoch datapoints \\
 ---    &   n2     & number of second-epoch datapoints \\
 ---    &   n\_del & number of deleted datapoints \\
pix     &   max\_res & largest residual in both proper-motion fits\\ 
 pix    &   X      & X-coordinate in ACS/WFC pixels aligned with RA\tablenotemark{a} \\
 pix    &   Y      & Y-coordinate in ACS/WFC pixels aligned with Decl \\
 deg    &   RAdeg  & Right Ascension, decimal degrees (J2000) \\
 deg    &   DEdeg  & Declination, decimal degrees (J2000) \\
---     &   flag   & reliability flag of selected proper motions \\ 
\enddata
\tablenotetext{a}{Direction of X-coordinate is opposite to RA}
\tablecomments{Table 1 is available in its entirety in machine-readable
format.}
\end{deluxetable*}

A low number of reference stars may produce very poor proper motions,
especially near the edges of ACS/WFC frames, while all other
astrometric parameters appear to be reasonable. Such cases would yield
locally large proper motions. We conjecture that, within a small
spatial spot, there should always be at least one object with a small
proper motion. If this is not true, then the measured proper motion
might be biased.  Therefore, within 25$\arcsec$ around each fast
moving source, we examined the proper motions of surrounding sources.
If there is at least one source with relatively small motion
($\mu\leq2$~mas~yr$^{-1}$) consistent with cluster membership, we
assumed that the proper motion of a fast moving source
($\mu\geq5$~mas~yr$^{-1}$) is reliable. Among the 141 such fast moving
sources, there are 26 possibly-unreliable sources, all marked in the
catalog.

In order to have an external check for our larger proper motions
($\mu\geq4$~mas~yr$^{-1}$; there are a total of 469 such objects), we
used the astrometric information provided by \citet{jo88}, {\it Gaia}
DR\,2, and \citet{ki19}. This exercise resulted in 64 sources with
their proper motions confirmed at least by one independent dataset, 15
sources with potentially unreliable proper motion, and 34 sources for
which we found our measurement discrepant and, thus, marked
accordingly. We also found 40 cases when the external source itself
(including {\it Gaia} DR\,2) has an unreliable proper motion.  None of
such objects with suspicious proper motion is used in the following
analysis.

\section{Discussion and Applications}\label{discuss}

The primary objective of this project is to provide proper motions and
the related tangential velocities for a large sample of low-mass stars
in the ONC. The stars with measured proper motions cover an irregular
area of $\sim$550 arcmin$^2$ with the approximate center at
R.A.$=5^{h}35^{m}$ and decl.$=-5^{\circ}$$27\arcmin$. We adopted the
\citet{da14} center of the Trapezium as the center of the ONC
(Figure~\ref{fig:spat_distr}).  While the longest spatial extension of
our field outwards from the Trapezium is $\sim$20\arcmin, in the
East-North direction it is significantly shorter (only
$\sim$10\arcmin) and, thus, misses a substantial fraction of the ONC.
 
In order to mitigate detrimental effects of a small number of datapoints, we
selected only those sources that have at least two datapoints at each epoch
and have proper-motion errors smaller than 0.4 mas~yr$^{-1}$. There are a total
of 1,379 sources with such properties and {\it only this sample is used in our
analysis}.  It should be noted that we probe a limited mass range of the ONC:
from $\sim$$0.4 M_\sun$ down to $\sim$$0.02 M_\sun$, as estimated from
the common sources with derived stellar parameters \citep[Table~3,][]{da12}.
Most likely, the majority of unused sources have reliable proper motions  with
the exception of those which have $\mu\geq$5 mas~yr$^{-1}$. 

\subsection{Internal Velocity Dispersion}\label{ivd}

An obvious application of our relative proper-motion catalog is a new
estimate of internal velocity dispersion (IVD) for the ONC.
Given a significant uncertainty in the distance of the ONC based on {\it Gaia}
DR\,2 data: $403$$\pm$7~pc \citep{ku19} {\it vs.} 389$\pm$3~pc \citep{ko18},
we consider only tangential velocities expressed in mas~yr$^{-1}$.
In a typical star cluster, the first task would be to obtain membership
probabilities. However, in the context of this study, the ONC is not an
ordinary cluster. 
It is located approximately towards the Galactic anticenter and significantly
away from the Galactic equator. The light from background stars is essentially
blocked out by a dense molecular cloud. Effectively, these factors make
the presence of field stars minimal. Next, due to the young age of the ONC,
$\sim$$2.5$ Myr \citep{da14}, the color-magnitude diagram of low mass
pre-main-sequence stars is just a fuzzy band \citep{hi97,da12}, exacerbated
by a strong differential reddening \citep{da12}. Therefore, a membership
selection based on the color-magnitude diagram alone could partially miss
a sizeable number of genuine cluster members.  Finally, the presence
of confirmed runaway objects in the Orion Kleinmann-Low Nebula 
\citep{ro17,lu17}, and exotic accreting young stellar objects \citep{ma13},
indicate that the usage of traditional kinematic memberships
\citep[e.g.,][]{jo88} may eliminate such extremely interesting sources. 
In the presence of these distinctive circumstances, we
assume that all sources in our area of the ONC are members, as long as some
other parameter (e.g., parallax) is not in obvious conflict with the cluster
membership. This argument was also applied by \citet{hi97} to infer the total
mass of the ONC.

\subsubsection{New Estimates of IVD}\label{new_ivd}

The spatial distribution of our ONC objects is heterogenous, due to the
layout of {\it HST} observations (Figure~\ref{fig:spat_distr}) and
the light-blocking effects
by nebulosity. Therefore, we defined two characteristic spatial structures:
a circular area around the Trapezium and a South-North prolongation
crossing the Trapezium. The radius of the former is $3\farcm45$, equivalent
to three times the core radius from the King model by \citet{hi98,ku19}.
The South-North prolongation is limited to the sources with R.A. $>$ $83\fdg75$.
Both features are marked in Figure~\ref{fig:best_spat_distr}.
The next step is to set reasonable limits in the vector-point diagram of proper motions. Faster moving sources are a mix of foreground
stars, ejected objects from the ONC, and poorly-measured proper motions.
None of them should be contributing to the calculated IVD.
Guided by the prior estimates of IVD, a relatively clean sample of the ONC
members can be delineated by an ellipse with the major and minor axes of
9.0 and 6.3 mas~yr$^{-1}$, centered at $\mu_x=+0.04$ and
$\mu_y=+0.08$~mas~yr$^{-1}$ (Figure~\ref{fig:best_vpd}).
This ellipse follows the velocity distribution along
 $\sim$3.5$\times\sigma_{v}$, where a subscript $v$ indicates the
observed 2-D proper-motion dispersion. 

The internal velocity dispersion is calculated using the standard 
formulation \citep[e.g.,][]{jo88}. For an estimate of the formal error,
we used the method described in \citet{va10}, which takes into account
individual uncertainties of proper motions.
An additional contribution by likely systematic errors is not yet feasible
to quantify.

\begin{deluxetable*}{cccclc}
 \tablecolumns{6}
 \tablewidth{0pt}
 \tablecaption{Samples of Internal Velocity Dispersion in mas~yr$^{-1}$\label{tab:ivd_plat}}
 \tablehead{
    \colhead{Sample \#} &
    \colhead{Area$^\mathrm{a}$}   &
    \colhead{Magnitude range}   &
    \colhead{$\sigma_{v,x}$}      &
    \colhead{$\sigma_{v,y}$}    &
    \colhead{N${_{\rm stars}}$}
}
\startdata
1 &   All & All  & $0.94\pm0.02$ & $1.25\pm0.03$ & 1197\\
2 & $r<3\farcm45$ & $<20.0$ & $0.91\pm0.04$ & $1.22\pm0.05$ & 293 \\
3 & $r<3\farcm45$ & $>20.0$ & $0.84\pm0.06$ & $1.20\pm0.08$ & 104 \\
4 & $r>3\farcm45$ & $<20.0$ & $0.91\pm0.03$ & $1.20\pm0.04$ & 361 \\
5 & $r>3\farcm45$ & $>20.0$ & $1.01\pm0.04$ & $1.32\pm0.05$ & 439 \\
6 & East, $r>3\farcm45$ & $<20.0$ & $0.89\pm0.04$ & $1.16\pm0.05$ & 300 \\
7 & East, $r>3\farcm45$ & $>20.0$ & $0.97\pm0.04$ & $1.33\pm0.06$ & 325 \\
\enddata
\tablenotetext{a}{Consult Figure~\ref{fig:best_spat_distr} for the area
conventions.
Meaning of East is a selection of sources eastward from R.A.$=83\fdg75$.}

\end{deluxetable*}

\begin{deluxetable*}{lclcr}
 \tablecolumns{5}
 \tablewidth{0pt}
 \tablecaption{Internal Velocity Dispersion in mas~yr$^{-1}$\label{tab:ivd_all}}
 \tablehead{
    \colhead{Source}            &
    \colhead{$\sigma_{v,x}$}      &
    \colhead{$\sigma_{v,y}$}    &
    \colhead{Spatial extent}    &
    \colhead{N${_{\rm stars}}$}
}
\startdata
van Altena et al. (1988) & $0.76\pm0.09$ & $0.83\pm0.10$$^\mathrm{a}$ & radius=30$\arcmin$ & 49 \\
Jones \& Walker (1988) & $0.91\pm0.05$ & $1.18\pm0.04$ & 25$\arcmin\times$30$\arcmin$ & 693 \\
Dzib et al. (2017) & $1.08\pm0.07$ & $1.27\pm0.15$$^\mathrm{b}$ & 2$\arcmin\times$2$\arcmin$ & 79 \\
Kuhn et al. (2019)  & $0.73\pm0.05$ & $1.12\pm0.10$$^\mathrm{c}$ & 20$\arcmin\times$18$\arcmin$ & 48 \\
Kim et al. (2019) & $0.83\pm0.02$ & $1.12\pm0.03$  & 6$\arcmin\times$6$\arcmin$ & 701 \\
This study & $0.89\pm0.03$ & $1.21\pm0.04$$^\mathrm{d}$ & radius=$3\farcm45$ & 397 \\
\enddata
\tablenotetext{a}{All stars brighter than $V=12.5$ mag.}
\tablenotetext{b}{From interferometric radio observations.}
\tablenotetext{c}{Positional angle of 5\arcdeg applied. In their notation, the pc2 axis is the $X$-axis (R.A.) as in all other studies.}
\tablenotetext{d}{Approximately within the spatial area of Kim et al. (2019); a significantly lower number in our sample is due to the selection of reliable proper-motion measurements.}
\end{deluxetable*}

The calculated internal velocity dispersion and its formal error for various
samples of the ONC members are given in Table~\ref{tab:ivd_plat}.
In order to uncover a potential dependence of the IVD on stellar mass,
we partitioned all subsamples at $m_{\rm F775W}=20.0$ mag. Note that near the
Trapezium (Figure~\ref{fig:best_spat_distr}, circular area) there is a clear
shortage of very-low-mass objects. In part, this is a consequence of extreme
contamination by various artifacts caused by the bright Trapezium stars.
Variations of the IVD across the FOV and for different stellar masses
are of the same order as our estimates of the IVD error.
The angular asymmetry of IVD is present at all stellar masses and locations.
An elevated $\sigma_{v,x}$ for Sample~5 ($r>3\farcm45$ and $m_{\rm F775W}>20.0$)
is puzzling. The only different parameter between the ``brighter" and ``fainter"
samples is a significantly higher number of fainter sources at decl. 
$<-5\fdg45$.  If we exclude these sources, then our global IVD estimate
of the ONC is $\sigma_{v,x}=0.94$ and $\sigma_{v,y}=1.25$ mas~yr$^{-1}$.

\subsubsection{Complete List of Measured IVDs}\label{comp_ivd}

\begin{figure}
%\resizebox{\hsize}{!}
\centering
{\includegraphics[width=5.7cm]{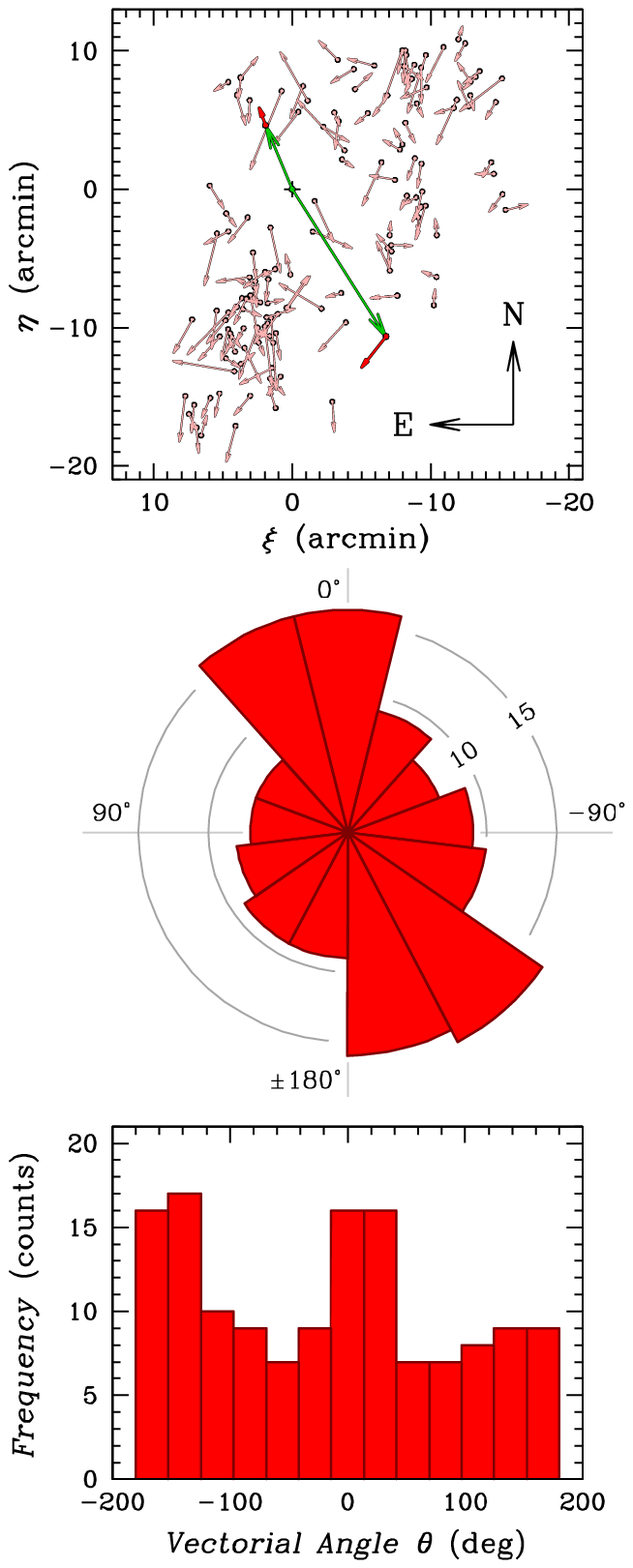}}
\caption{From a spatial map of tangential velocities of the ONC to
one-dimensional histogram of vectorial angles. 
   Upper panel: a spatial distribution of 141 fast-moving sources and their
proper motions in the tangential plane with zeropoint at the location of
Trapezium, marked by a small cross. The length of arrows is proportional
to the size of proper motions. Vectorial angle between the position vector
and the arrow of a proper motion is a key parameter to identify potential
runaways.  Position vectors (in green) are drawn to the location
of two sources: a Northern source has a very small vectorial angle
$\Theta$=$-2\fdg3$ and is a likely runaway candidate, while a Southern
source has $\Theta$=$-71\arcdeg$ and is a passing-by field star.
   Middle panel: polar histogram of vectorial angles for all sources in the
upper panel. The size of bins is 27\fdg7 and the upper bin is
centered on $\Theta$=$0\arcdeg$. The radial coordinate indicates the number
of sources in each bin.
Bottom panel: same as in the middle panel but transformed into the
one-dimensional histogram of vectorial angles. Only the central
bin at $\Theta$=$0\arcdeg$ contains likely runaways. The presence of other
peaks is discussed in Section~\ref{run_ori}.
}
\label{fig:vectorial_angles}
\end{figure}

\begin{figure*}
%\resizebox{\hsize}{!}
\centering
{\includegraphics[width=14cm]{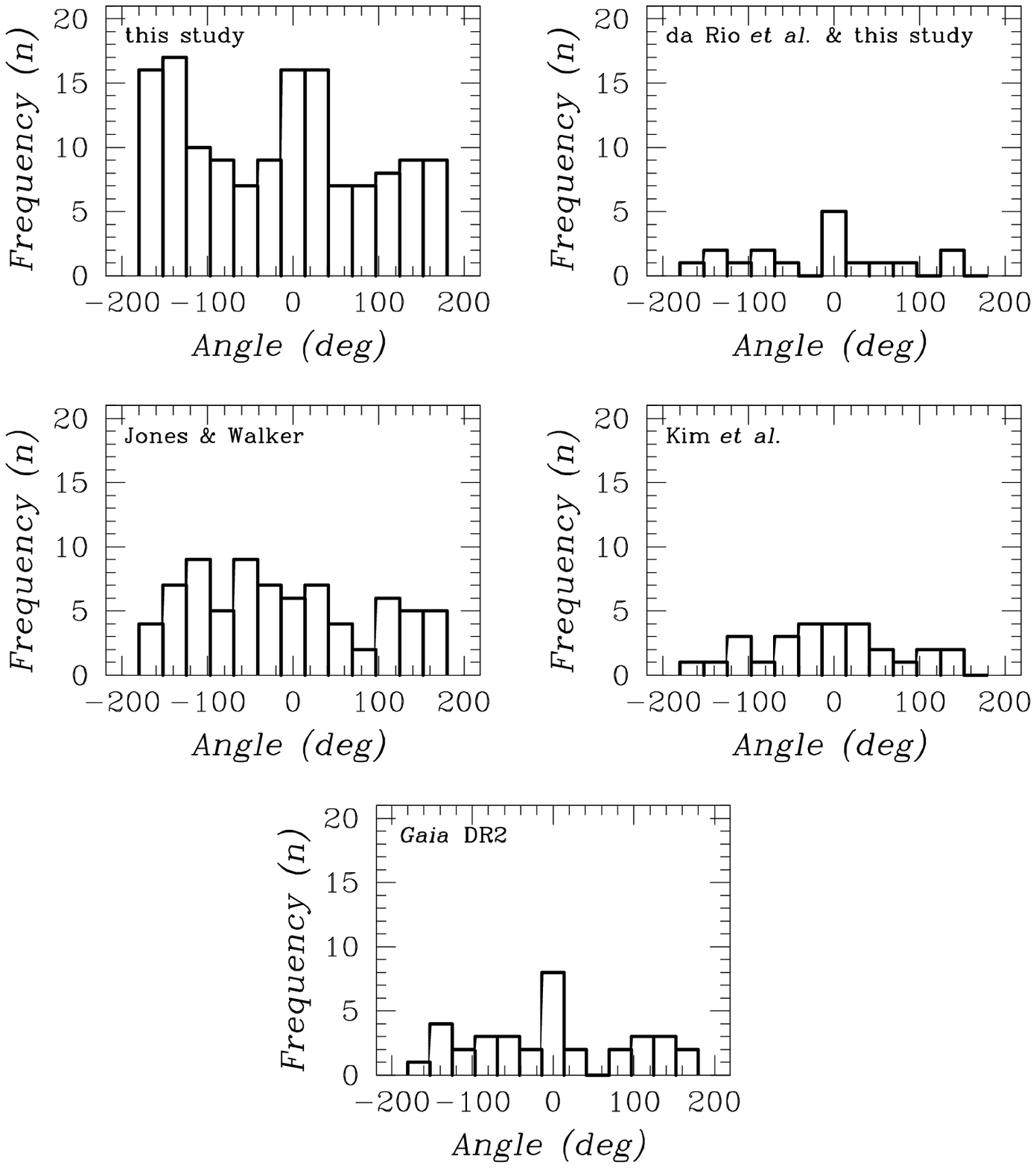}}
\caption{One-dimensional histograms of vectorial angles for sources with
larger proper-motions. The source (authors) of proper motions
is indicated in each panel. A significant excess of sources at a zero angle
in three of the five histograms indicates the presence of runaway sources.
A detailed analysis is given in Section~\ref{runaway}.
}
\label{fig:angle_5hist}
\end{figure*}

Until recently, only two estimates of the IVD were available \citep{jo88,va88}.
Table~\ref{tab:ivd_all} provides a significant addition, including our effort.
It is not trivial to compare these estimates because of differences in
the  spatial and brightness coverage as well as due to the variety of
reduction techniques for ground and space observations.
It is expected that the final {\it Gaia} data will provide the best estimate
of the IVD for sources more massive than  $\sim$$0.3 M_\sun$. In turn,
the {\it HST} is still the prime instrument for measuring proper motions 
of low-mass stars and brown dwarfs in the ONC. Within the errors, our study
and that by \citet{jo88} provide nearly identical IVDs. Note that, 
in Table~\ref{tab:ivd_all}, our estimate of the IVD
is limited to the inner area around the Trapezium. The only outlier
appears to be the $\sigma_{v,y}$ value obtained by \citet{va88}. However,
internal kinematics of the brighter stars might be different than that of
the low-mass stars.

There are two studies, by \citet{ku19} and \citet{ki19}, which provide
a somewhat lower IVD than that of the other studies, including ours. 
It appears odd that largery the same {\it HST} observations used
by us and \citet{ki19} can produce a $\sim$$7\%$ smaller IVD in the latter
study. On the other hand, \citet{ki19} expanded the time baseline to 20
years (vs. the 11 years of this study) by including additional {\it HST}
observations taken with other cameras through a variety of filters.
Such favorable conditions -- a longer time baseline and additional epochs --
must significantly lower the proper-motion errors. However,
Figure~\ref{fig:kim_plat} shows the opposite; overall, for the 572 stars in
common, the proper-motion errors by \citet{ki19}  are significantly larger.
If we calculate the IVD following the \citet{jo88} formulation and applying
the same spatial and magnitude cut-offs as in Section~\ref{new_ivd} but using
the \citet{ki19} data and their errors,
then the resulting formal dispersions along the $X$ and $Y$ axes, 
respectively, are 0.78 and 1.09 mas~yr$^{-1}$.  Assuming zero
errors, the same dispersions are 0.85 and 1.15 mas~yr$^{-1}$. 
This exercise demonstrates that larger proper-motion errors tend to lower the
calculated IVD. Apart from the unusal distribution of proper-motion errors,
there are no obvious clues to explain the likely-underestimated IVD by
\citet{ki19}. The IVD estimate based on the
{\it Gaia} DR\,2 \citep{ku19} is consistent with other studies in
the $Y$-direction but it is significantly smaller in the $X$-direction.
For the ONC area, this is consistent with significally larger proper-motion
errors along the RA in {\it Gaia} DR\,2.

While working on potential runaway stars in the ONC, we conjectured that
consideration should be given to another scenario. It is related to the
North-South kinematic ``stream'' moving relatively fast across the ONC (see
following Section~\ref{runaway}). Most likely this stream is part of young
stellar objects in the Orion\,A molecular cloud \citep{me12}.
The proper motions of brighter ONC stars by \citet{va88} might be the least
polluted by this stream and, in turn, their IVD estimate is more reliable given
a similar size along {\it both} axes (Table~\ref{tab:ivd_all}).
If indeed all other samples of ONC members are significantly contaminated,
then one must somehow identify the true ONC members, that is, the objects
dynamically associated with the Trapezium. Proper motions alone, no matter
how accurate, cannot provide a clean sample of bona-fide cluster members
due to the intrinsically large IVD. Our proper motions
indicate that possible  kinematic differences in the tangential plane between
the young stellar objects (YSO) of Orion\,A and Trapezium are less
than 0.1 mas~yr$^{-1}$ and mainly
 in the $X$-direction. However, a study of the Orion\,A 3D-shape \citep{gr18}
indicates that the bulk of its YSOs might be in the foreground
of the ONC. A similar argument is provided in Appendix\,A by \citet{ku19}.
We believe that the final {\it Gaia} parallaxes will decisively separate
these two populations.

\subsection{ONC Candidate Runaways}\label{runaway}

The current status of potential runaway sources from the ONC is discussed in
\citet{mk19}. They appear to cover the entire range of spectral types 
from O9.5\,V down to late M stars.   
Following earlier attempts to identify runaway objects from the ONC, such
as \citet{ta04,po05,ki19,mk19}, we explored various sets of proper motions
for such objects. Considering that the ONC contains not only the fully-formed
stars but also  proplyds and protostars (e.g., the Becklin-Neugebauer object),
here we adopt the short-hand term ``runaway'' to characterize all sources and
objects in the process of being ejected from the ONC regardless of escapee's
velocity. One way to identify such
objects in the ONC is to use the virial theorem and the estimate of the mean-square
escape velocity at $\sim$3.1 mas~yr$^{-1}$ (5.9~km~s$^{-1}$, assuming
a distance of 400~pc), beyond which an object is classified
as a runaway \citep{ki19}.
We complemented this approach with an additional constraint
by adding the vectorial angle between the proper motion with
respect to the direction outward from the Trapezium \citep[e.g.,][]{pl18}.
Given the importance of the Trapezium in various dynamical processes
\citep[e.g.,][]{al17,po16}, it should also be a principal (but not the only
one) engine for the production of runaways.
We note that the spatial density of potential runaways is proportional
to the inverse of squared distance from the Trapezium. In other words,
the highest chance of finding a runaway by proper motions is near
the cluster.  
In the following, we discuss the status of runaways in five sets of proper
motions.

\subsubsection{{\it HST} Treasury Programs of ONC (This Study)}\label{run_ori}

We considered all sources with total proper motions
larger than 5 mas~yr$^{-1}$ and a total error less than 0.4 mas~yr$^{-1}$.
There are a total of 141 such sources (Figure~\ref{fig:vectorial_angles}).
This figure shows how the vectorial angles are obtained and used to
identify candidate runaways.
In order to quantify the distribution of vectorial angles, we chose 13
bins with the width of $27\arcdeg7$. The bin size is optimized considering
the errors of proper motions and the scatter of vectorial angles for some
known candidate runaways. Figure~\ref{fig:vectorial_angles} indicates
a concentration of vectorial angles in the bin centered on
$\Theta$=$0\arcdeg$; that is, where the potential Trapezium runaways
are expected.  At all other angles the frequency of stars should be flat,
provided that the distribution of field-star proper motions is random. 
We note that a relatively large bin-size may enable to detect
potential runways originating also from other massive stars and stellar
systems of the ONC.

We noticed, however, an unusual excess of sources in the bins at 
$\Theta$=$-166\arcdeg$, $-138\arcdeg$, and $+28\arcdeg$. 
The majority of sources from the bin $\Theta$=$+28\arcdeg$ are located in
the South-East quadrant 
(Figures~\ref{fig:vectorial_angles},\ref{fig:angle_5hist}).
In fact, these sources appear to have the same kinematic pattern as
the sources in the bins at $\Theta$=$-166\arcdeg$, $-138\arcdeg$ that are by
$\sim$$180\arcdeg$ apart and located 
in the North-West quadrant. This pattern we interpret
as the presence of a stellar stream running approximately North-South.
The distribution of proper-motion vectors in polar coordinates of our 141
fast-moving sources indicates the dominant direction towards
$\varphi$=$162\arcdeg$ (from North to East) with the associated 
Gaussian FWHM=$68\arcdeg$.
These parameters very well describe the pattern visible in the histograms
( Figures~\ref{fig:vectorial_angles}, \ref{fig:angle_5hist}; left upper panel).
All four peaks are mainly due to the field stars streaming across the ONC.
Our kinematic data alone cannot be used to identify potential runaways because
of dominating background stars. Even more, an equal number of sources in
the bins at $\Theta$=$0\arcdeg$ and $+28\arcdeg$ effectively rules out
the presence of runaways in our dataset. As a result, we single out only
source \#7320 (Table~\ref{tab:run_oricat}), that has its
proper-motion error just a little bit exceeding the adopted precision
threshold but formally has the smallest vectorial angle among all
fast-moving stars.  

\begin{table*}
\centering
\small{
\centering
\begin{tabular}{lcccccc}
\multicolumn{7}{c}{\textsc{\bf Table 4.} Candidate low-mass runaways from ONC\label{tab:run_oricat}}\\
\hline\hline
Ident & F775W & X & Y & Total PM & Angle & d\\
      & mag & arcmin & arcmin &  mas~yr$^{-1}$ & deg & \arcsec \\
\hline
6177 & 17.603 & \phs1.9207 & \phn\phs4.6279 & \phn6.81 &  $-$2.3 & \phn12\\
7320$^\mathrm{a}$ & 23.369 & \phs4.0906 &     $-$17.0977 &    13.50 &  $-$1.0 & \phn18\\
7495 & 21.721 & \phs4.6063 &  \phn$-$3.0388 & \phn9.79 &  $-$5.6 & \phn33\\
7863 & 22.077 & \phs5.9075 &     $-$15.0817 & \phn7.74 &  $-$2.0 & \phn33\\
\hline
\multicolumn{7}{l}{(a) Not a counterpart in the ONC-member list by \citet{da12}.}\\
\multicolumn{7}{l}{Last column shows the extrapolated nearest angular distance from}\\
\multicolumn{7}{l}{ the Trapezium (impact parameter).}\\
\end{tabular}}
\end{table*}

One of the most reliable censuses of the low-mass ONC members down to
$\sim$$0.02 M_\sun$ was produced by \citet{da12}. This study used empirical
relations to determine effective temperature $T_{\rm eff}$ and total
extinction $A_V$ from optical colors. These parameters were used to 
construct the Hertzsprung-Russell diagram and, then, to identify likely
cluster members.  We cross-correlated the list of these cluster members
with our proper-motion catalog and examined the vectorial angles
adopting the same parameters as above. There are a total of 819 common
sources, including 17 common fast-moving objects
(Figure~\ref{fig:angle_5hist}; upper right panel). Statistically, the
distribution of vectorial angles implies a single field object and four
candidate runaways. We selected three sources as our best sample of
potential runaways within the mass range of 0.11-0.35\,$M_\sun$,
 all having the impact parameter less than
50$\arcsec$ (Table~\ref{tab:run_oricat}), well-aware that the same sources
contributed to the histogram of all fast-moving objects 
(Figure~\ref{fig:angle_5hist} upper left panel). 
The estimated standard error of vectorial angles for these candidate
runaways is $\sim$0\fdg5 and their total proper motions
span from 6.8 to 9.8 mas~yr$^{-1}$, equivalent of 13.3 to 19.2 km~s$^{-1}$.

Summarizing our contribution to the subject of ONC runaways, we acknowledge
that it is ambiguous. For example, if the authors of the low-mass 
ONC member list \citep{da12} would have had access to our proper motions,
then all our fast-moving sources could be classed as background objects
\citep[see;][]{da10}. This conundrum necessitates a complex study of each
candidate runaway; high priority should be given to obtaining a set of
key astrophysical parameters. If they match those of the bona fide cluster
members then the likelihood of being a former cluster member would be
irrefutable. 

We noticed a trapezium-like configuration among the fast moving low-mass
field objects comprising \#1115, 1127, 1167, and 1169. Their mutual
separation is about 15$\arcsec$ and three of them are comoving. The kinematics
of these objects is not compatible with the ONC membership nor a runaway
status.  As expected, none of them is listed in \citet{da12}. Nonetheless, it
is an interesting case in terms of formation and stability of such systems.

Finally, we briefly explored whether, besides the Trapezium, some other
massive ONC members \citep[Table~4,][]{mu08} can produce additional candidate
runaways. Inside the circle ($r=8\arcmin$) centered onto
the Trapezium there are 5 such stars -- Brun~655, 682, 714, 747, and 760.
All of them produce vectorial-angle histograms similar to
Figure~\ref{fig:angle_5hist} (upper left panel) with a significant amount
in the bin $\Theta$=$0\arcdeg$ and a heavy spatial concentration in the
South-East quadrant. Similar to the case of Trapezium, it is not possible
to identify  reliable runaways. However, we note that Brun~747 is a massive
hierarchical triple system \citep{sh19} capable to eject a cluster member.

\subsubsection{\citet{jo88}}\label{run_jones}

A relatively deep survey of the ONC proper motions by \citet{jo88} can
in principle be used to search for runaways.
\citet{po05} suggested three runaways (also known to be
the proplyds) among this set of proper motions, a claim that was disproved
\citep{od05}.  We confirm that these objects do have small
proper motions and cannot be astrometric runaways. There are a few other
cases of inflated proper motions and/or underestimated membership probabilities
in the Jones \& Walker catalog \citep[see][]{hi97}, likely due to the
effects of partially resolved visual binaries. The distribution
of proper-motion vectorial angles of fast moving stars
(Figure~\ref{fig:angle_5hist}) is inconclusive about the presence of
potential runaways. Among the formal six potential runaways, source JW~45
appears to be the only candidate runaway confirmed by {\it Gaia} DR\,2.

\subsubsection{\citet{ki19}}\label{run_kim}

The analysis of proper motions by \citet{ki19} was one of the first attempts
to identify escaping cluster members among the high-velocity stars that
are habitually
classified as background sources. These authors adopted the angular escape
velocity at 3.1~mas~yr$^{-1}$, above which a star can be considered escaping
from the ONC. We found that, in the group of high-significance escapers
(ESC\,1), sources Kim 209 and 232 have significantly smaller proper motions
in our catalog. If we calculate the vectorial angle, then only Kim 572, 611
and 713 appear to originate near off the Trapezium. None of them is in our
catalog.  The histogram of vectorial angles (Figure~\ref{fig:angle_5hist})
indicates 1-2 potential runaways. Source Kim 713 is the Becklin-Neugebauer
object BN -- a well-known escapee from the ONC area \citep{ro17}.

\subsubsection{{\it Gaia} DR\,2}\label{run_gaia}

There is a very limited overlap
between the {\it HST} candidate low-mass runaway sources and the potentially
brighter runaways in {\it Gaia} DR\,2. However, runaways are expected over
a wide range of masses, not only in the low-mass domain.

In order to mimic our FOV, we selected DR\,2 stars within 20$\arcmin$ off the
Trapezium with a total proper-motion larger than 4~mas~yr$^{-1}$ and the
total error $\sigma_\mu$$<$0.55~mas~yr$^{-1}$. The absolute proper
motions from DR\,2 were translated into the system of our relative proper
motions.  Initially, we considered all {\it Gaia} stars with parallaxes
1.9$<$$\varpi$$<$3.1~mas. 
Within these specifications, there are 35 fast moving stars.
This range of $\varpi$ still comprises a large distance range, between
320 and 530~pc. Probing a clean sample of the ONC members along
the line-of-sight direction would require the parallax errors no larger than
$\sim$0.03~mas, while the current sample has an average parallax
error of 0.1~mas.  

The histogram of vectorial angles for our selection of DR\,2 stars
shows a convincing peak in the zero bin (Figure~\ref{fig:angle_5hist}), and
thus reinforces our conclusions on the ONC runaways.
Statistically, it indicates a total of 6 runaways. However, a lesson learned
from our {\it HST} proper motions (Section~\ref{run_ori}) compels us to narrow
the range of well-measured parallaxes within 1$\sigma$ around the average
$\varpi$=2.50~mas. This cutoff eliminates nearly all potential runaway
candidates
in the controversial South-East quadrant and produces 4 likely runaway
candidates (Table~\ref{tab:run_gaia} and Figure~\ref{fig:runaway_distr}). 
There is one visual binary: sources \#1 and \#2 are at a spatial separation of
17\farcs7 but they cannot be a physical pair due to the $\sim$$10\sigma_{\mu}$
difference in proper motions. This pair is also discussed by \citet{mk19}.
In fact, Table~\ref{tab:run_gaia} contains only one new candidate runaway
which is not listed  by \citet{mk19}. This is source \#1 = JW~45 (see
Section~\ref{run_jones}) = Parenago~1540 \citep{ma88}. It is
a pre-main-sequence double-lined spectroscopic binary. \citet{ma88} are
the first to propose that Parenago~1540 might be a runaway from the Trapezium.
The {\it Gaia} DR\,2 data clearly support this proposition.

\begin{table*}
\centering
\small{
\centering
\begin{tabular}{lccccccccc}
\multicolumn{10}{c}{\textsc{\bf Table 5.} Likely Escaped Stars from ONC in {\it Gaia} DR\,2\label{tab:run_gaia}}\\
\hline\hline
Ident & RA (DR\,2) & Dec (DR\,2) & $\varpi$ & $G$ & $X$ & $Y$ & Total PM & Angle & d \\
 & deg & deg & mas & mag & arcmin & arcmin & mas~yr$^{-1}$ & deg & \arcsec \\
\hline
1 & 83.66566645 & $-$5.40711238 & 2.495$\pm0.043$ & 10.818  & \phn$-$9.0844 & \phn$-$1.1545 &  \phn5.24 & \phn$-$3.3 & \phn32 \\
2$^\mathrm{a}$ &  83.61391001 & $-$5.40619446 & 2.546$\pm0.041$ & 11.609 & $-$12.1760 & \phn$-$1.1003 &  \phn8.55 & \phn$-$1.8 & \phn23\\
3$^\mathrm{a}$ &  83.60906346 & $-$5.40527690 & 2.519$\pm0.044$ & 12.713 & $-$12.4656 & \phn$-$1.0454 &  \phn7.25 &  \phn$-$1.2 & \phn16 \\
4$^\mathrm{a}$ & 83.76792173 & $-$5.13684069 & 2.431$\pm0.037$ & 10.130 & \phn$-$2.9777 & \phs15.0629 &   \phn7.18 &  \phn\phs2.8 & \phs46\\
5$^\mathrm{b}$ & 83.80879167 & $-$5.37296388 & \nodata & \nodata & \phn$-$0.5351 &  \phn\phs0.8956 & 13.50 & \phn$-$6.0 & \phn\phn7\\
\hline
\multicolumn{10}{l}{(a) Identified by \citet{mk19} as a source consistent with its origin from the Trapezium area.}\\
\multicolumn{10}{l}{(b) The Orion\,BN object is not observed by {\it Gaia}. 
Presented parameters are based on the Very Large Array}\\
\multicolumn{10}{l}{    observations \citep{ro17}.}\\
\end{tabular}}
\end{table*}

\subsubsection{\citet{mk19}}\label{mk_gaia}

\begin{figure}
{\includegraphics[width=\columnwidth]{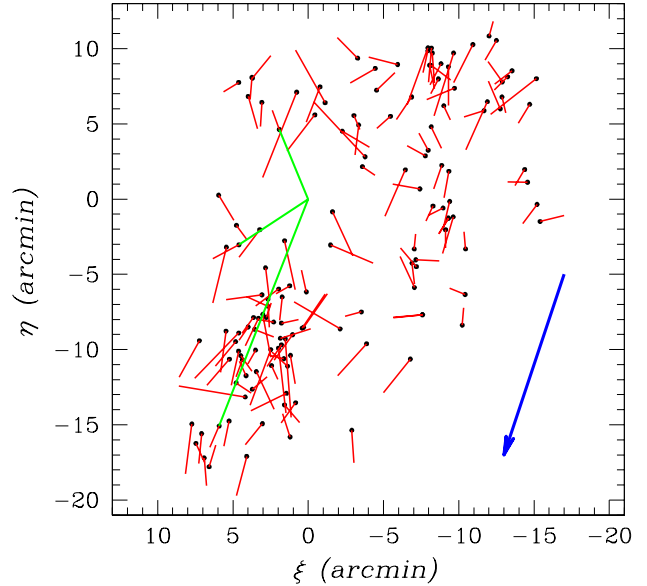}}
\caption{Spatial distribution of our faster-moving sources. Proper-motion
vectors are shown by red line-segments at a scale of 5 mas~yr$^{-1}$ per
$1\arcmin$. The total proper motions range from 5 to 30 mas~yr$^{-1}$.
Three green lines with their origin at the Trapezium indicate new
low-mass candidate runaways from our HST Treasury programs of the ONC.
Blue arrow shows the dominant direction of Orion's A faster-moving young 
stellar objects.
}
\label{fig:fast_distr}
\end{figure}

\begin{figure}
%\resizebox{\hsize}{!}
{\includegraphics[width=\columnwidth]{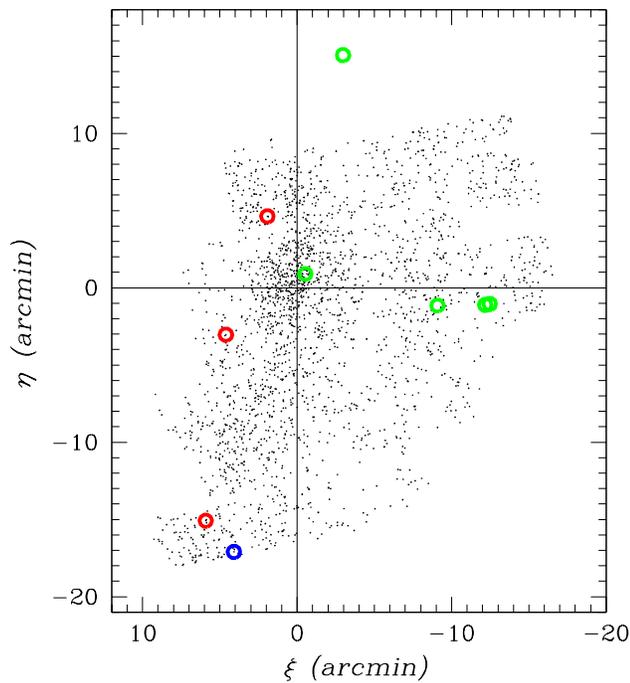}}
\caption{Spatial distribution of candidate ONC runaways. Background objects are
the same as in Figure~\ref{fig:spat_distr} but shown in tangential coordinates
$\xi$, $\eta$. Runaways are divided in three parts:
1) confirmed by {\it HST} proper motions the likely \citet{da12} photometric
low-mass members of the ONC (red circles);
2) an additional candidate from {\it HST} proper motions (blue);
3) selected candidates from the {\it Gaia} DR\,2 catalog (green).
A nearest runaway to the Trapezium is the BN object, not detected
by {\it HST} nor {\it Gaia} but also marked green.
}
\label{fig:runaway_distr}
\end{figure}

A new study of runaway 
young stars near the ONC \citep{mk19} is conceptually closest to our approach. These authors used a sample of {\it Gaia} DR\,2
data within the radius of 2\arcdeg around the Trapezium. The most interesting
part of this study is their proposed  sources originating from the Trapezium.
There are 9 such sources. Among them, three (sources~$a$, $b$, and $g$) are in common with our selection of runaways in {\it Gaia} DR\,2 (Table~\ref{tab:run_gaia}).
Source~$e$ has a large impact parameter at
$\sim$9\arcmin, indecisive  {\it Gaia} astrometry, and is located in
the controversial South-East quadrant. Similarly, source~$c$ has poor
{\it Gaia} astrometry and a larger  impact parameter, incompatible with
the origin in the Trapezium. In turn, source~$d$ has a $4.5\sigma$ parallax
offset from the mean ONC parallax provided by \citet{ku19}, thus indicating
a background object.  If we exclude these three sources, then these
authors have contributed a total of 6 new candidate runaways out to
$\sim$1$\arcdeg$ around the Trapezium. {A caveat to this list is a strong
spatial alignment along the North-South direction,
which mimics the distribution of
fast-moving field YSOs in our data (Figure~\ref{fig:fast_distr})}. 

\subsubsection{Census of ONC Runaways}\label{run_status}

Summarizing, our {\it HST} ACS/WFC study alone produced 3 new ONC candidate
runaways in the magnitude range  17.6$<$$m_{\rm F775W}$$<$22.1~mag. 
The status of an additional source 7320 is not clear
(see Section~\ref{run_ori}),
therefore it is omitted in this census.  In addition, we propose one new
{\it Gaia} DR\,2 candidate runaway at $G$=10.8 mag. Formally, the enigmatic
Orion BN object can be classed as an ONC runaway, however, it seems to be
a member of a multiple system in the Orion Kleinmann-Low Nebula
\citep[e.g.,][]{lu17,ro17}. This scenario is partially weakened by the failure
to detect the water vapor in close proximity to BN \citep{in18}.
Given the unclear status of BN, we ignore it in our census.

Considering an additional 6 runaways identified by \citet{mk19}, the current
census of ONC runaways includes a total of 10 candidates. However, considering
the caveats to various datasets, our census may still contain $\sim$50\%
impostors, By the same token, we may have overlooked some additional
genuine runaways. 
The range of equivalent tangential velocities of all these candidate runaways
is between 10 and 19 km~s$^{-1}$. It is surprising not to find any faster
moving objects. This is telling us that there might be just a single scenario
to producing them. 

The average velocity of this sample is $\sim$16 km~s$^{-1}$, equivalent
to 8.0 mas~yr$^{-1}$. If we assume an angular distance of 20$\arcmin$
from the Trapezium and allow our sample to move at this velocity, then
it takes $\sim$150,000 years to cover this distance. 
Assuming that out to
20$\arcmin$ there are $\sim$5 genuine runaways and the age of ONC is
$\sim$2 Myr, this would result in $\sim$70 lost sources with a mass less
than 1\,$M_{\sun}$. Considering the estimated total number of cluster 
members at $\sim$2800 \citep{hi98}, this appears to be a minor $\sim$3\%
loss of the ONC members and thus would be a negligible correction
to the mass function of the ONC.

\section{Conclusions}\label{conclu}

We derived a new proper motion catalog for low-mass stars and other sources
in the area of the ONC using {\it HST} ACS/WFC observations over an 11-year
time span. The resulting catalog of relative proper motions contains 2,454
objects in the magnitude range of $14.2<m_{\rm F775W}<24.7$. A subset of
sources with high-precision proper motions is used to estimate the internal
velocity dispersion of the ONC. The same subset also reveals  the presence
of three new low-mass candidate runaway sources, a result that is supported by
extensive analysis of the candidate ONC runaways with a likely origin
from the Trapezium.

In summary, our main findings are as follows:

\begin{enumerate}

\item We provide a detailed account how to obtain reliable proper motions
with {\it HST} in the area of the ONC that is otherwise notoriously difficult
for astrometric studies.

\item We estimated the internal velocity dispersion of the ONC in two
ranges of magnitude and for some distinctive spatial features in the area of
the ONC.  At a level of the achieved precision, the IVD appears to be nearly
constant over the field-of-view.

\item The calculated internal velocity dispersion in the area around
the Trapezium ($\sigma_{v,x}=0.94$ and $\sigma_{v,y}=1.25$ mas~yr$^{-1}$)
matches closely the estimate by \citet{jo88}, but disagrees with those of
\citet{ki19} at the 7\% level. At this time, none of the available IVD
estimates seems to be more reliable than the others. The {\it Gaia} DR\,2
data cannot provide a reliable list of genuine ONC members decisively
separated from the young stellar objects in Orion\,A. We suspect that
the high value of $\sigma_{v,y}$ is due to our current inability to obtain
a clean sample of the ONC members.

\item Using the {\it HST} ACS/WFC data alone, we identified 3 new 
faster-moving low-mass sources with their likely origin in the Trapezium.
In addition, one more candidate runaway is identified in {\it Gaia} DR\,2.
Altogether, the current census of the ONC candidate runaways is 10 sources
covering a wide range of apparent magnitudes. Our analysis indicates that
the number of bona fide runaways might be lower and their impact onto
the observed ONC mass function appears to be insignificant.  
 
\end{enumerate}

\acknowledgements  The authors gratefully acknowledge grant support
for programs GO-10246 and GO-13826, provided by NASA through grants from
the Space Telescope Science Institute, which is operated  by the Association of
Universities for Research in Astronomy, Inc., under NASA contract
NAS~5-26555. We thank Terrence Girard for his expert opinion on IVD and
Marina Kounkel for sharing the main data on their selection of runaways.
We also thank an anonymous referee for the insightful review of our manuscript.
This work has made use of data from the European Space Agency
(ESA) mission {\it Gaia} (\url{https://www.cosmos.esa.int/gaia}),
processed by the {\it Gaia} Data Processing and Analysis Consortium (DPAC,
\url{https://www.cosmos.esa.int/web/gaia/dpac/consortium}). Funding for the DPAC
has been provided by national institutions, in particular the institutions
participating in the {\it Gaia} Multilateral Agreement.

{\it Facilities:} \facility{Hubble Space Telescope}
%%%%%%%%%%%%%%%%%%%%%%%%%%%%%%%%%%%%%%%%%%%%%%%%%

\end{document}